\documentclass[twocolumn, prl, superscriptaddress]{revtex4-1}
\usepackage{amssymb}
\usepackage{amsmath}
\usepackage{epsfig}
\usepackage{color}
\usepackage{graphics, graphicx}
\usepackage{bbold}
\usepackage{psfrag}
\usepackage{mathcomp}
\usepackage{subfigure}
\usepackage{verbatim}
\usepackage{color}
\usepackage[colorlinks,citecolor=blue]{hyperref}
\usepackage{braket}

\begin{document}
\title{Boson-Anyon-Fermion Mapping and Anyon Construction in One Dimension}
\author{Haitian Wang}
\affiliation{Beijing National Laboratory for Condensed Matter Physics, Institute of Physics, Chinese Academy of Sciences, Beijing 100190, China}
\affiliation{School of Physical Sciences, University of Chinese Academy of Sciences, Beijing 100049, China}
\author{Yu Chen}
\affiliation{Beijing National Laboratory for Condensed Matter Physics, Institute of Physics, Chinese Academy of Sciences, Beijing 100190, China}
\affiliation{School of Physical Sciences, University of Chinese Academy of Sciences, Beijing 100049, China}
\author{Xiaoling Cui}
\email{xlcui@iphy.ac.cn}
\affiliation{Beijing National Laboratory for Condensed Matter Physics, Institute of Physics, Chinese Academy of Sciences, Beijing 100190, China}
\date{\today}

\begin{abstract}

We establish an exact mapping  between identical particles in one dimension with arbitrary exchange statistics, including bosons, anyons and fermions, provided they share the same scattering length. This boson-anyon-fermion mapping facilitates the construction of anyons  from a linear superposition of spatially symmetric and anti-symmetric states. This scheme is general and has been demonstrated in a spin-1/2 Fermi gas, where both s- and p-wave bound states can be supported by manipulating spin channels. 
With a suitable symmetry-breaking field, these bound states are hybridized to form a fractional-wave molecule. The condensation of these molecules in a many-body system leads to anyonic superfluidity, characterized by fractional statistics upon spin exchange within a Cooper pair. 
These anyonic states can be detected through asymmetric momentum distributions for each spin with a chiral $k^{-3}$ tail.  Our results have demonstrated the inadequacy of contact interaction model for anyons in continuum and lattices, and meanwhile proposed a convenient route for engineering fractional phases  in the platform of ultracold atoms.
\end{abstract}

\maketitle

The statistics and interactions of quantum particles are often closely related, in that different exchange symmetries determine distinct scattering properties. For instance, identical bosons and fermions are dominated by s- and p-wave scatterings at low energy, respectively, resulting in completely different interaction forms. In one dimension (1D), the correlation between statistics and interaction is even pronounced because particles can only exchange via collision. This leads to many unique properties of 1D systems, and one such unique property is an exact mapping  between identical bosons and fermions with short-range interactions. The boson-fermion mapping was initially established  for hardcore bosons and non-interacting fermions\cite{Girardeau1}, and later extended to general couplings with reversed role of coupling strengths\cite{Cheon}. Importantly, it reveals the equivalence of bosons and fermions in 1D under certain interaction conditions, providing a power tool for understanding universal quantum physics in two seemingly distinct systems.

Beyond bosons and fermions, quantum particles can also manifest as anyons with fractional statistics, characterized  by a  fractional phase $\alpha$  that interpolates between $0$ (boson) and $\pi$ (fermion) upon their position exchange. 
Over the past decades, anyon physics has generated significant interests ranging from fractional quantum Hall effects in electron materials\cite{Stern,Wilczek} to  non-abelian braiding for topological quantum computations\cite{Simon}. 
Recently, anyonic statistics has been experimentally realized in lattice systems using electric circuit\cite{Zhang1} or ultracold atoms with density-dependent hoppings\cite{Kwan}. Theoretically,  static and dynamical properties of 1D interacting anyons have been extensively studied in both continuum\cite{Kundu,Guan,Patu,Chen2,Girardeau2, Pelster2, Campo, Piroli}
 and lattices\cite{Keilmann,Pelster1,Pelster2,Santos,Longhi,Chen1,Wang,Zhang, Gorshkov, Wang2}. 

Despite these developments, the exploration of anyon physics still encounters fundamental challenges. First, most existing studies have assumed a contact ($\delta$-function) interaction for anyons based on Kundu's model\cite{Kundu}, which is identical to that for bosons. Given the distinct statistics between bosons and anyons,  this assumption is quite questionable. For example, fermions experience a completely different (p-wave) interaction as compared to (s-wave) bosons  due to their different statistics. As of now, it remains unclear how to properly describe the interaction of anyons, which should cover both bosons and fermions by varying phase $\alpha$.  With a proper description, one might be able to identify a more general relation between 1D systems with arbitrary statistics, but not limited to bosons and fermions\cite{Girardeau1, Cheon}. Secondly, the experimentally realized anyons in lattices turn out to experience the same on-site interaction and on-site permutation relation as bosons\cite{Kwan}. Such anyon-Hubbard model, sharing the same spirit as Kundu's model, fails to capture the crucial effect of fractional statistics to short-range collision when two anyons approach each other. 
Therefore, in the experimental side it is imperative to engineer anyon physics without lattices, such as in continuum or a single trap, where the interplay of fractional statistics and short-range interaction can be more transparently visualized. Unfortunately, a feasible scheme to engineer anyon without lattices is still lacking at the moment.

This work aims to address these problems.  We have identified a unified description of short-range interactions for all 1D systems with different statistics, by utilizing the short-range boundary condition for pairwise collision on one side. Such boundary condition is fully parametrized by scattering length ($l$), while the statistics ($\alpha$) just determine the asymptotic wavefunction on the other side of the collision. Based on this description, we have established an exact mapping between 1D systems with arbitrary statistics, including bosons, anyons and fermions, termed the boson-anyon-fermion (BAF) mapping. 
It tells that all these systems  can be mapped to each other in both energy and real-space wavefunction as long as they have the same $l$.  
The BAF mapping suggests a convenient new way to construct anyons, i.e., by directly superposing   spatially symmetric (boson-like) and anti-symmetric (fermion-like) states at the basic two-body level. The scheme is general and has been demonstrated in a spin-1/2 Fermi gas, which can support both s- and p-wave bound states in different spin channels.
Under a proper symmetry breaking field,  these bound states are hybridized  to form  fractional-wave molecules.  The condensation of these molecules further  leads to  anyonic superfluidity, characterized by fractional statistics upon spin exchange in a Cooper pair. These anyonic states can be detected through asymmetric  momentum distributions for each spin with a chiral $k^{-3}$ tail. Finally, we  discuss possible ultracold systems to realize our scheme.

We start by considering two identical particles ($x_1,x_2$) moving in 1D with relative and center-of-mass (CoM) coordinates  $x=x_2-x_1$ and $R=(x_1+x_2)/2$. Their wavefunction obeys
\begin{equation}
\Psi(R,-x)=e^{i\alpha}\Psi(R,x), \ \ \ \ \ (x>0)  \label{statistics}
\end{equation}
where $\alpha$ determines quantum statistics and has the period $2\pi$. In this work we will take half of a period $\alpha\in[0,\pi]$, since $\alpha$ and $-\alpha$ are just related by the reflection symmetry  
$x\leftrightarrow-x$. Boson, fermion and anyon then respectively correspond to $\alpha=0$, $\pi$ and $\in(0,\pi)$. 

\begin{figure}[t]
\includegraphics[width=9cm]{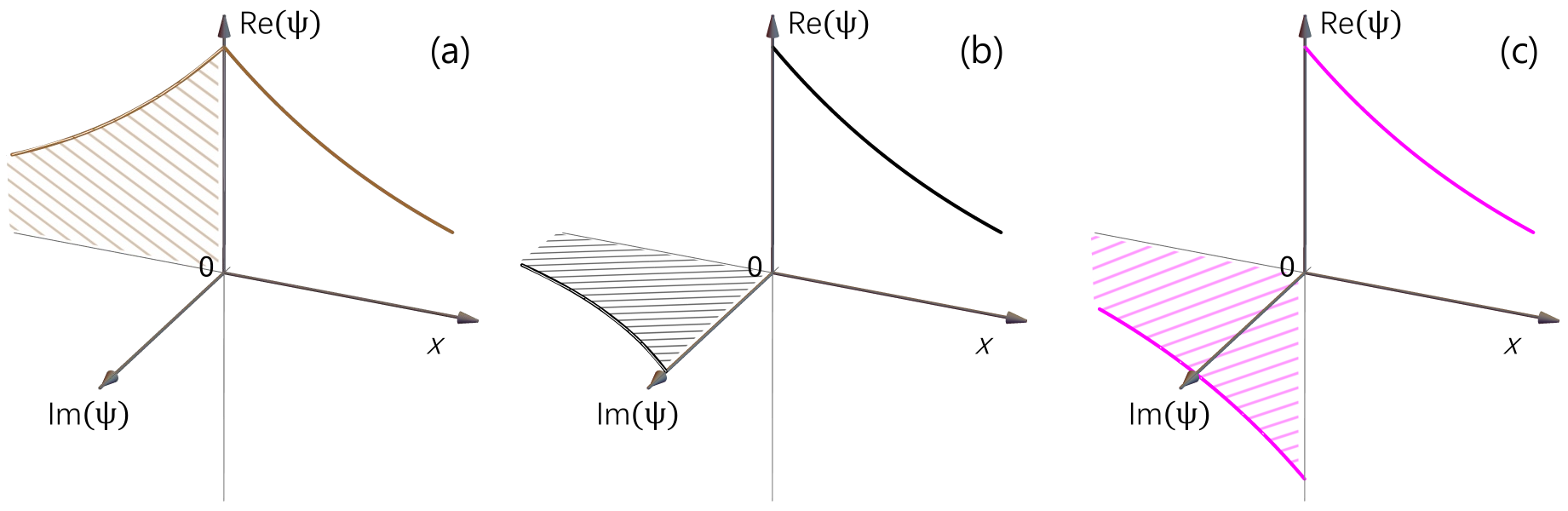}
\caption{(Color online) Typical two-body wavefunctions $\Psi$ as functions of relative coordinate $x\equiv x_2-x_1$ in 1D for bosons (a), anyons with $\alpha=\pi/2$ (b) and fermions (c). For all cases, $\Psi$ at $x>0$ are taken to be real and share the same function.   
The shaded area aims to highlight $\Psi$ at $x<0$. 
} \label{fig_wf}
\end{figure}

When turn on short-range interactions, $\Psi$ and/or its derivatives will develop certain discontinuities when two particles approach each other ($x\rightarrow 0^{\pm}$). At $x\rightarrow 0^+$, $\Psi$ asymptotically behaves as  
\begin{equation}
\Psi(R,x)\rightarrow f(R) (x-l), \ \ \ \ \ (x\rightarrow 0^+)  \label{bc}
\end{equation}
where $l$ is defined as the scattering length\cite{footnote}. 
According to Eq.(\ref{statistics}),  $\Psi$ at $x\rightarrow 0^-$ automatically follows 
\begin{equation}
\Psi(R,x)\rightarrow f(R) e^{i\alpha}  (-x-l). \ \ \ \ \ (x\rightarrow 0^-) \label{bc2}
\end{equation}
Obviously, different statistics ($\alpha$) lead to different behaviors of $\Psi$ at short range, as shown in Fig.\ref{fig_wf}. For bosons ($\alpha=0$), $\Psi$ is continuous and well-defined at $x\rightarrow 0^{\pm}$ (Fig.\ref{fig_wf}(a)), and therefore the interaction of bosons simply follows a $\delta$-function potential in s-wave channel:
\begin{equation}
U_s=g_s\delta(x). \label{Us}
\end{equation}
For fermions ($\alpha=\pi$), $\Psi$ is no longer continuous at $x\rightarrow 0^{\pm}$ but $\Psi'\equiv \partial\Psi/\partial x$ is (Fig.\ref{fig_wf}(c)), so their interaction can only occur in  p-wave channel:
\begin{equation}
U_p=g_p\partial_x\delta(x)\partial_x.  \label{Up}
\end{equation}
For anyons, however, neither $\Psi$ nor $\Psi'$ is well-defined at $x=0^{\pm}$ (Fig.\ref{fig_wf}(c)), and therefore their mutual interaction cannot solely follow $U_s$ or $U_p$. This is why the anyon model with a $\delta$-function potential (identical to $U_s$) in previous studies is problematic. 
 
As shown in Fig.\ref{fig_wf}, the interaction effect has actually been fully incorporated in the asymptotic behavior of $\Psi$ on one side of collision ($x>0$), while $\Psi$ on the other side ($x<0$) is just given by exchange statistics ($\alpha$). It follows that one can simply utilize the short-range boundary condition on one side ($x>0$) to describe the interaction effect. For bosons, this boundary condition is equivalent to applying $U_s$ (Eq.(\ref{Us})), and for fermions it can also be related to $U_p$ (Eq.(\ref{Up})) given that the bare coupling $g_p$ is renormalized\cite{cui}. Here we generalize this approach  to arbitrary statistics: for a general 1D system of $N$ particles with wavefunction $\Psi(x_1,...x_N)$, we write the short-range boundary condition as 
\begin{equation}
\lim_{x\equiv x_j-x_i\rightarrow 0^+} \left( \frac{1}{l} + \partial_x\right) \Psi(x_1,x_2,...x_N)=0. \label{BC}
\end{equation}
The exchange statistics for a given $\alpha$ requires
\begin{equation}
\Psi(x_1,...x_j,...x_i,...x_N)=e^{i\alpha w}\Psi(x_1,...x_i,...x_j,...x_N), \label{exchange}
\end{equation}
where $w=\sum_{k=i+1}^j \epsilon(x_k-x_i)-\sum_{k=i+1}^{j-1} \epsilon(x_k-x_j)$, with $\epsilon(x)=1$ for $x>0$ and $-1$ for $x<0$. Note that $\epsilon(x)$ is {\it not} defined at $x=0$ due to the discontinuity of $\Psi$ at this singular point (Fig.\ref{fig_wf}(b)). This is different from  previous studies of interacting anyons based on Kundu's model\cite{Kundu}.

Eq.(\ref{BC}) provides a unified description of short-range interaction for 1D systems with arbitrary statistics, where the scattering length $l$ serves as the unique physical quantity to characterize interaction strength. Based on this description, we demonstrate here an exact mapping between 1D systems with different statistics ($\alpha$), termed the {\it boson-anyon-fermion (BAF) mapping: all $\alpha$-systems can be mapped to each other in both energy and real-space wavefunction as long as they share  the same $l$}. When applied to two limiting cases $\alpha=0,\pi$, it reproduces the boson-fermion mapping\cite{Girardeau1, Cheon}, where the reversed role of coupling strength is equivalent to requiring the same $l$ for both systems.

A rigorous way to prove BAF mapping in  homogeneous case is through the Bethe-ansatz solutions. As shown in the supplementary material\cite{supple}, all $\alpha$-systems share the same Bethe-ansatz equation at a given $l$, thereby ensuring the same quasi-momentum distribution and the same energy. This is in contrast to previous studies of anyons with $\delta$-function potential, where the coupling constant in Bethe-ansatz equation is modified from that of bosons by a factor related to $\alpha$\cite{Guan,Patu,Chen2,Kundu, Piroli}. This modification can be attributed to the artificial definition of anyon wavefunction at  singular point $x_i=x_j$.  In comparison, here we apply Eq.(\ref{BC}) to pairs of anyons coming close to each other but not exactly staying at the same site.

The BAF mapping can also apply to inhomogeneous systems with external potentials. A general proof is as follows. First, let us denote $\psi(x_1,x_2,...x_N)$ as the wavefunction of $N$ particles aligning in the region $x_1<x_2...<x_N$. We can prove that all $\alpha$-systems share the same $\psi$ (up to a global phase) and the same energy $E$. It is because for all $x_i\neq x_{i+1}$ the system is simply governed by the non-interacting Hamiltonian, which is the same for all $\alpha$-systems; whenever two neighboring particles come close ($x_{i+1}\rightarrow x_i+0^+$), all systems satisfy the same boundary condition (Eq.(\ref{BC}))  with the same $l$. Therefore all these systems have the same $\psi$ and $E$, regardless of the value of $\alpha$. Secondly, for any other region, the wavefunction can be related to $\psi$  via  Eq.(\ref{exchange}), where $\alpha$ just determines their relative phase  but does not change the eigen-energy of the system. 

By revealing the intimate relation between anyonic and bosonic/fermionic 1D systems,  the BAF mapping provides us an important insight for constructing anyons.
At the basic two-body level, this can be done by directly superposing the spatially symmetric (s-wave, boson-like) and anti-symmetric (p-wave, fermion-like) wavefunctions under the same $l$, denoted by $\phi_s(x)$ and $\phi_p(x)=\epsilon(x)\phi_s(x)$ respectively, through:  
\begin{equation}
\phi_{\rm any}(x)=\phi_s(x)-i\tan\frac{\alpha}{2}\phi_p(x). \label{superpose}
\end{equation}
By such superposition, $\phi_{\rm any}$ satisfies anyonic statistics $\phi_{\rm any}(-x)=e^{i\alpha}\phi_{\rm any}(x)$ $(x>0)$. 
Eq.(\ref{superpose}) implies that the interaction of anyons is also the superposition of  $U_s$ and $U_p$, where $U_s$ ($U_p$) just takes effect on $\phi_s$ ($\phi_p$) but not the other part. This is drastically different from Kundu's model with only contact ($U_s$) interaction, which is equivalent to a bosonic model with complicated double-$\delta$ potentials\cite{Kundu,Pelster2}. Generalized to lattice systems, the anyon interaction should involve both on-site (s-wave) and nearest-neighbor (p-wave) interactions, rather than simply the on-site one as in  anyon-Hubbard model\cite{Kwan, Keilmann,Pelster1,Pelster2,Santos,Longhi,Wang, Gorshkov}. Next, we will show anyons can be engineered in physical systems with realistic two-body potentials. The key challenge is to find both s- and p-wave states in a single system and meanwhile hybridize them with a pure   imaginary coefficient ($\sim i \tan(\alpha/2)$). 

\begin{figure}[t]
\includegraphics[width=9cm]{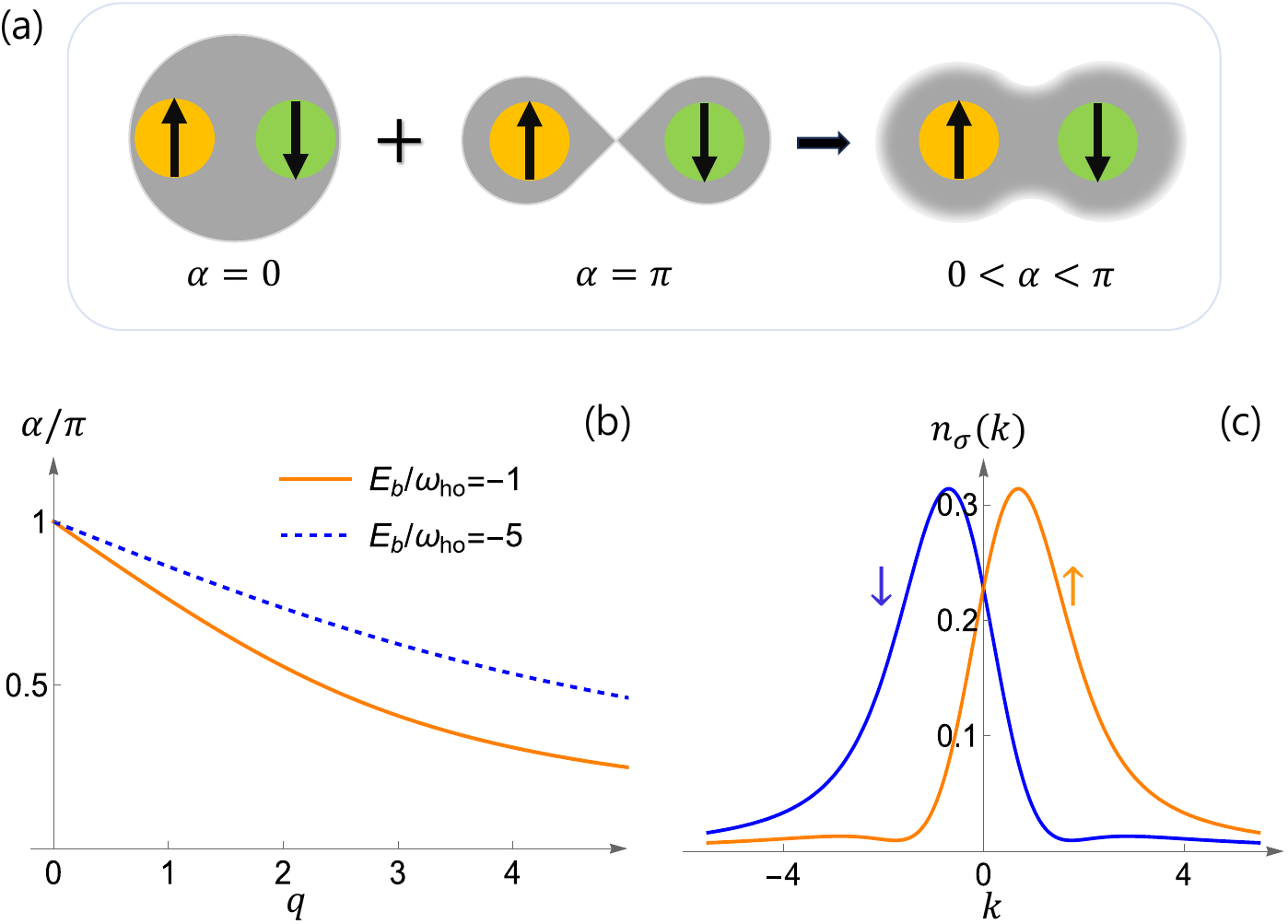}
\caption{(Color online) (a) Schematics of the hybridization between spatially symmetric ($\alpha=0$)  and anti-symmetric ($\alpha=\pi$) two-body states to achieve a fractional-wave molecule with fractional statistics ($0<\alpha<\pi$). (b) Fractional phase $\alpha$ as a function of $q$ (in unit of $1/l_{ho}$) for two fermions  in a harmonic trap with equal s- and p-wave binding energies $E_b/\omega_{ho}=-5,-1$. (c) Momentum distribution of $\uparrow$ and $\downarrow$ spins for fractional-wave molecule at $\alpha=\pi/2$ and  $E_b/\omega_{ho}=-1$.  Here $n_{\sigma}$ and $k$ are, respectively, in units of $l_{ho}$ and $1/l_{ho}$.
} \label{fig_anyon}
\end{figure}

Here we propose the spin-1/2 ($\uparrow,\downarrow$) system, 
where both spatially symmetric and anti-symmetric states can be supported by manipulating the associated spin channels. Take spin-1/2 fermions for example, the s-wave interaction between $\uparrow$ and $\downarrow$ can support a spatially symmetric state in spin-singlet channel:
\begin{equation}
\Psi_s(R,x)=\Phi_0(R)\phi_s(x)|\uparrow_1\downarrow_2-\downarrow_1\uparrow_2\rangle/\sqrt{2};
\end{equation}
and the p-wave interaction supports a spatially anti-symmetric state in spin-triplet channel:
\begin{equation}
\Psi_p(R,x)=\Phi_0(R)\phi_p(x)|\uparrow_1\downarrow_2+\downarrow_1\uparrow_2\rangle/\sqrt{2},
\end{equation}
with $\Phi_0$ the ground state of CoM motion. Under a suitable symmetry breaking field, these two states can hybridize to result in anyonic states with fractional statistics, as schematically shown in Fig.\ref{fig_anyon} (a). 

To be concrete, we consider two fermions ($\uparrow,\downarrow$) in a harmonic trap, with frequency $\omega_{ho}$ and typical length $l_{ho}=(m\omega_{ho})^{-1/2}$. Further, we take the symmetry breaking field as a weak spin-orbit coupling\cite{soc_review}:
\begin{equation}
V_{\rm soc}=\Omega \sum_i \big[ e^{iqx_i}\sigma^{+}_{i}+ e^{-iqx_i}\sigma^{-}_{i}\big], \label{soc}
\end{equation}
where $q$ is the transferred momentum and $\sigma^{\pm}=\sigma_x\pm i\sigma_y$ ($\sigma_{x,y,z}$ are Pauli matrices).  
Given the same positive s- and p-wave scattering length $l_s=l_p\equiv l>0$, both s- and p-wave molecules can be supported with the same binding energy $E_b$. 
Based on the second-order perturbation theory, $V_{soc}$ induces virtual excitations of these molecules ($\Psi_s$ and $\Psi_p$) to other spin states ($|\uparrow\uparrow\rangle$ and $|\downarrow\downarrow\rangle$) at higher levels of relative and CoM motions\cite{supple}. This leads to an effective Hamiltonian $H_{\rm eff}=v{\cal M}v^T$ with $v=(\Psi_s,\Psi_p)$ and 
\begin{equation}
{\cal M}=-\frac{\Omega^2}{\omega_{ho}} \left(\begin{array}{cc}A & iC \\-iC & B\end{array}\right). \label{M}
\end{equation}
Here $A,B,C$ are all real numbers that solely depend on $E_b/\omega_{ho}$ (or $l/l_{ho}$) and $ql_{ho}$. Crucially, $V_{\rm soc}$ gives rise to a purely imaginary off-diagonal coupling $\sim \pm iC$, which is necessary for anyon construction  in Eq.(\ref{superpose}). 
Finally, we arrive at the  ground state wavefunction:
\begin{equation}
\Psi_G=\Phi_0(R)\Big[(\phi_s-i\tan\frac{\alpha}{2}\phi_p)|\uparrow_1\downarrow_2\rangle-(\phi_s+i\tan\frac{\alpha}{2}\phi_p)|\downarrow_1\uparrow_2\rangle\Big], \label{anyon_wf}
\end{equation} 
with 
\begin{equation}
\tan\frac{\alpha}{2}\equiv\frac{1}{C}\left(\sqrt{\left(\frac{A-B}{2} \right)^2+C^2}-\frac{A-B}{2} \right).
\end{equation}
Remarkably, $\Psi_G$ describes a fractional-wave molecule, in that the position exchange of $\uparrow$ and $\downarrow$ exactly gives rise to a fractional phase  $\alpha$\cite{supple}. 

Fig.\ref{fig_anyon}(b) shows how $\alpha$ varies with $ql_{ho}$ for different molecule binding energies. Take $E_b/\omega_{ho}=-1$ for example, we can see a highly tunable $\alpha$ from $\pi$ to $\sim \pi/4$ as increasing $ql_{ho}$ from $0$ to $5$. Fig.\ref{fig_anyon}(c) further shows spin-dependent momentum distribution, $n_{\sigma}(k)$, for fractional-wave molecule at $\alpha=\pi/2$ and  $E_b/\omega_{ho}=-1$. These distributions exhibit strong asymmetry, i.e., $n_{\sigma}(k)\neq n_{\sigma}(-k)$, and we have checked that such asymmetry preserves for all fractional $\alpha\in (0,\pi)$ but not for purely s-wave ($\alpha=0$) and p-wave ($\alpha=\pi$) cases. 
Similar asymmetries in momentum distribution or transport dynamics of anyons were also revealed previously\cite{Chen2,Gorshkov,Wang2,Kwan, Chen1, Wang, Pelster1,Campo, Piroli}. In our case, despite of the loss of symmetry for each spin, a spin-momentum combined symmetry is preserved  as $n_{\uparrow}(k)= n_{\downarrow}(-k)$. Such combined symmetry is protected by the reflection symmetry of CoM motion, i.e., $\Phi_0(R)=\Phi_0(-R)$\cite{supple}.  

The physical origin of  asymmetric $n_{\sigma}(k)$ can be seen more clearly from the free space case, where the wavefunctions of s- and p-wave molecules simply follow $\phi_s=e^{-\kappa |x|}$ and $\phi_p=\epsilon(x)e^{-\kappa |x|}$ ($\kappa=(m|E_b|)^{-1/2}$), and their Fourier transformations are (up to a common factor)
\begin{equation}
\phi_s(k)=\frac{\kappa}{k^2+\kappa^2},\ \ \ \phi_p(k)=\frac{-ik}{k^2+\kappa^2} \label{phi_k}
\end{equation}
The fractional-wave molecule in Eq.(\ref{anyon_wf}) can be created by 
\begin{equation}
d^{\dag}=\sum_k \left[ \phi_s(k)-i\tan\frac{\alpha}{2}\phi_p(k) \right] c_{-k\uparrow}^{\dag}c_{k\downarrow}^{\dag}, \label{dimer}
\end{equation} 
where $c_{k\sigma}^{\dag}$ is to create a spin-$\sigma$ fermion at momentum $k$. From Eqs.(\ref{phi_k},\ref{dimer}), we  can obtain 
\begin{equation}
n_{\sigma}(k)\sim \left(\frac{\kappa+\epsilon_{\sigma}\tan(\alpha/2)k}{k^2+\kappa^2} \right)^2.\ \ \ \ (\epsilon_{\uparrow}=1,\ \epsilon_{\downarrow}=-1)  \label{nk}
\end{equation} 
Now it is clear that asymmetric momentum distributions of anyons exactly come from the  superposition  of s- and p-wave states (with imaginary superposition coefficient), and therefore such asymmetry can be seen as a characteristic feature of fractional statistics.  At high $k$, such asymmetry is reflected in the tail $\sim k^{-3}$, which has an opposite sign for different spins and thus is chiral. Note that this should be distinguished from $\sim k^{-3}$ tail in previous studies that are either due to center-of-mass excitations\cite{Cui_Dong} or appear in off-diagonal distributions\cite{Qin}.

In a many-body system, these fractional-wave molecules can condense and give rise to superfluidity. Given $d^{\dag}$ in Eq.(\ref{dimer}) to create one such molecule\cite{footnote2}, we write down a  coherent state $\sim e^{\lambda d^{\dag}}$ for its condensation, which is equivalent to  a BCS-type wavefunction
\begin{equation}
|\Psi\rangle_{\rm ASF}=\prod_k \Big(1+\lambda\left[ \phi_s(k)-i\tan\frac{\alpha}{2}\phi_p(k) \right] c_{k\uparrow}^{\dag}c_{-k\downarrow}^{\dag}\Big). \label{ASF}
\end{equation}
Remarkably, Eq.(\ref{ASF}) represents an anyonic superfluidity (ASF) featuring fractional statistics upon spin exchange within a Cooper pair. Similar to  fractional-wave molecule,  ASF exhibits  asymmetric momentum distributions with chiral $k^{-3}$ tail and spin-momentum combined symmetry.

The anyonic states could be realized in two-component Fermi gases with coexistent s- and p-wave interactions. For instance,  a $^6$Li Fermi gas with hyperfine states $|F=1/2,m_F=1/2\rangle$ and $|F=3/2,m_F=-3/2\rangle$  undergoes a broad p-wave Feshbach resonance with low inelastic collision loss near $B\sim 225$G\cite{Luo}, and meanwhile it has a  large s-wave scattering length $a_s\sim 1000a_0$ ($a_0$ is the Bohr radius). Under confinement-induced resonances\cite{Olshanii, Blume}, this system may achieve the same s- and p-wave scattering length in quasi-1D. Note that this condition is relaxed to the same $E_b$ in the presence of effective range. 
In such a quasi-1D geometry, FWSF could be supported at low temperature when the phase fluctuations are suppressed\cite{Gora}. 
Another potential system is alkali-earth atoms with ground state $^{1}S_0$ and long-lived  $^{3}P_0$ states, which can experience s- and p-wave interactions in different orbital and spin channels and the interaction strengths can be conveniently tuned by magnetic fields or confinements\cite{Zhai_review}. The application of SOC  has also been very successful in ultracold gases of both alkali\cite{soc_review} and alkali-earth\cite{Ye_soc} atoms. 

We emphasize that  the anyon construction based on Eq.(\ref{superpose}) is very general and not limited to the specific case of fermions with positive $l$. For instance, it can also be implemented in two-species bosons, or in the hard-core limit with $l_s=l_p=0$. For the latter case, all spin states (singlet or triplet)  have the same energy due to spin-charge separation, giving a natural situation for the coexistence of s- and p-wave states in Eq.(\ref{superpose}). More details of constructing anyons in this case can be found in \cite{supple}. 
Moreover, SOC in Eq.(\ref{soc}) can also be replaced by other fields, as long as they produce a purely imaginary off-diagonal coupling as in Eq.(\ref{M}). 

In summary, we have established a general mapping between short-range interacting 1D systems with arbitrary statistics, encompassing  bosons, anyons and fermions. This BAF mapping  highlights  the scattering length as the unique physical quantity  to characterize interaction effect. Inspired by this mapping, we can construct anyons without resorting to lattices, i.e., by a proper  superposition of boson- and fermion-like states. For a demonstration, we have successfully constructed  fractional-wave molecule and anyonic superfluidity in a spin-1/2 Fermi gas.  These  states feature fractional statistics upon spin exchange within a basic two-body unit (molecule or Cooper pair), and such statistics can be visualized by  asymmetric spin distributions in momentum space. All these results could be detected in ultracold experiments of alkali or alkali-earth atoms.

Finally, we remark that the BAF mapping and its resultant anyon construction  are unique to 1D systems. For higher dimensions such as 2D or 3D, s- and p-wave states no longer have a good mapping as their short-range singularities are qualitatively different. For instance, in 3D the relative wavefunction in $l$-wave channel follows $\psi_l(r)\sim r^{-(2l+1)}$ at short distance $r$, and therefore it is impossible to achieve anyons by superposing different partial waves.   In this regard, the 1D geometry provide a rare and surprisingly simple situation to simulate intriguing quantum physics with fractional statistics.

\bigskip

{\bf Acknowledgement.} We thank Jiaming Li and Peng Zhang for useful discussions on potential experimental setups. 
This work is supported by the National Natural Science Foundation of China (12074419, 12134015), and the Strategic Priority Research Program of Chinese Academy of Sciences (XDB33000000).

\clearpage

\onecolumngrid
\vspace*{1cm}
\begin{center}
{\large\bfseries Supplementary Materials}
\end{center}
\setcounter{figure}{0}
\setcounter{equation}{0}
\renewcommand{\figurename}{Fig.}
\renewcommand{\thefigure}{S\arabic{figure}}
\renewcommand{\theequation}{S\arabic{equation}}

In this supplementary material, we provide more details on the exact solution of 1D anyons based on the short-range boundary condition, and on the construction of anyons in spin-1/2  systems. 

\section*{I.\ \ \ Exact solutions of 1D anyons}

The exact solutions of 1D anyons have been studied before based on the Kundu's model with $\delta$-function potential. Here we provide the exact solution based on the short-range boundary condition (Eq.(6) in the main text). 

According to the exchange symmetry (Eq.(7) in the main text), the anyon  wavefunction can be written as
\begin{align}
\Psi(x_{1}, \ldots, x_{N})&=\sum_{Q} \theta(x_{q_{N}}-x_{q_{N-1}}) \cdots \theta(x_{q_{2}}-x_{q_{1}})   \exp\biggl(i\frac{\alpha}{2}(\Lambda(x_{q_{1}},x_{q_{2}},\ldots,x_{q_{N}}))\biggr) \ \varphi(x_{q_{1}},x_{q_{2}},\ldots,x_{q_{N}})  \label{general_wf}
\end{align}
Here $\theta$ is the Heaviside step function; $\varphi(x_{q_1},x_{q_2},...,x_{q_N})$ is the wave function for the region $0\leq x_{q_1}\leq x_{q_2}\leq\cdots\leq x_{q_N}\leq L$ (with $L$ the system length); $Q=(q_1,...,q_N)$ presents a permutation of the position index of $N$ particles; $\Lambda=\sum_{j<k}^{N}\epsilon(x_j-x_k)$ with
\begin{equation}
\epsilon(x)=1\ (x>0);\ \ -1\ (x<0).
\end{equation}
Note that different from previous studies, here $\epsilon(x)$ is not defined at the singular point $x=0$. 

In the following, we will just consider a specific region $0\leq x_1< x_2<\cdots< x_{N}\leq L$. Moreover, we consider the system confined in a hard-wall potential, which satisfies the open boundary condition 
\begin{eqnarray}
\varphi(0,x_{2},...,x_{N})=0;\ \ \ \ \varphi(x_{1},x_{2},...,L)=0. \label{OBC}
\end{eqnarray}
The vanishing of $\varphi$ at the boundaries ($0$ and $L$) successfully avoids the complexities due to the (artificial) choice of phase difference between them.  

$\varphi$ can be expanded by plane-waves:
\begin{eqnarray}
\varphi(x_{1},x_{2},...,x_{N})=\sum_{P,\{\epsilon_j\}}\left[A_{P,\{\epsilon_j\}} \exp\left(i\sum_{j}\epsilon_j k_{p_j}x_{j} \right)\right]. \label{BA wavefunction}
\end{eqnarray}
Here $k_j(>0)$ ($j=1,...N$) presents the quasi-momentum, and $\epsilon_j=+1(-1)$ denotes the plane-wave of the $j$-th particle (with coordinate $x_j$) moving from left(right) to right(left) in coordinate space; $P=(p_1,p_2,\cdots,p_N)$ is a permutation of the momentum index, and $A_{P,\{\epsilon_j\}}\equiv A(k_{p_1},k_{p_2},\cdots,k_{p_N}; \epsilon_1,\epsilon_2,\cdots, \epsilon_N)$ is the superposition coefficients. 

Now we apply the short-range boundary condition, Eq.(6) in the main text, to obtain the relation between different coefficients. Given two neighboring coordinates $x_k$ and $x_{k+1}$, we take the terms when they are associated with momenta $\epsilon_ik_i$ and $\epsilon_jk_j$, and then arrive at the following equation by applying the short-range boundary condition at $x_{k+1}-x_k\rightarrow 0^+$:
\begin{eqnarray} 
&&\frac{i}{2}(\epsilon_jk_j-\epsilon_ik_i)\left[A(\ldots k_i,k_j\ldots;\ldots \epsilon_i,\epsilon_j\dots)-A(\ldots k_j,k_i\ldots;\ldots\epsilon_j,\epsilon_i\ldots)\right]  \nonumber\\
&=&-\frac{1}{l}\left[A(\ldots k_i,k_j\ldots;\ldots \epsilon_i,\epsilon_j\dots)+A(\ldots k_j,k_i\ldots;\ldots\epsilon_j,\epsilon_i\ldots)\right],
\end{eqnarray}
which gives
\begin{eqnarray}
  \frac{A(\ldots k_i,k_j\ldots;\ldots \epsilon_i,\epsilon_j\dots)}{A(\ldots k_j,k_i\ldots;\ldots\epsilon_j,\epsilon_i\ldots)}=\frac{\epsilon_{i}k_{i}-\epsilon_{j}k_{j}-2i/l}{\epsilon_{i}k_{i}-\epsilon_{j}k_{j}+2i/l}.
   \label{relation1}
\end{eqnarray}
The boundary condition in Eq.(\ref{OBC}) further requires
\begin{eqnarray}
&&  \frac{A(k_{P1},\cdots,k_{PN};\epsilon_1=1,\epsilon_2,\cdots,\epsilon_N)}{A(k_{P1},\cdots,k_{PN};\epsilon_1=-1,\epsilon_2,\cdots,\epsilon_N)}=-1;\nonumber\\
&& \frac{A(k_{P1},\cdots,k_{j};\epsilon_1,\cdots,\epsilon_N=1)}{A(k_{P1},\cdots,k_{j};\epsilon_1,\cdots,\epsilon_N=-1)}=-e^{-i 2k_jL}.
   \label{relation2}
\end{eqnarray}
Then we can use Eqs.(\ref{relation1},\ref{relation2}) to transmit coefficients $A(k_j,\cdots;\epsilon_1=1,\cdots) \to A(k_j,\cdots;\epsilon_1=-1,\cdots) \to A(\cdots,k_j;\cdots,\epsilon_N=-1) \to A(\cdots,k_j;\cdots,\epsilon_N=1)\to A(k_j,\cdots;\epsilon_1=1,\cdots)$, and finally we get:
\begin{equation}
\mathrm{e}^{\mathrm{i}2k_{j}L}=\prod_{i=1(\neq j)}^N\frac{k_j-k_i-2i/l}{k_j-k_i+2i/l}\frac{k_j+k_i-2i/l}{k_j+k_i+2i/l}. \label{BAE}
\end{equation}
One can see that (\ref{BAE}) is identical to the Bethe-ansatz equations for identical bosons\cite{textbook} and identical fermions\cite{Pan} under open boundary condition, given that these systems share the same scattering length $l$. Importantly,  the statistics ($\alpha$) just determines the general form of wavefunction in Eq.(\ref{general_wf}), but does not enter the derivation of Eq.(\ref{BAE}), which only relies on the short-range boundary condition in the region of $0\leq x_1< x_2<\cdots< x_{N}\leq L$. This proves the Boson-Anyon-Fermion mapping with exact Bethe-ansatz solutions. 

To  compare with previous studies, we note that a modified coupling constant (with an additional factor $\sim 1/\cos(\alpha/2)$) appears in the Bethe-ansatz equations therein. It is because these studies have adopted a contact interaction potential for anyons and also assumed a boson-like permutation relation when two anyons sit at the  same site ($x=x_i-x_j=0$). Here, we emphasize the invalidity of these boson-like treatments for anyon systems, given very different exchange symmetries and asymptotic wavefunctions at short range.  Instead,  to describe the interaction effect in 1D anyons,  we have utilized the short-range boundary condition just at one side of pairwise collision with $x\rightarrow 0^+$, but not exactly at $x=0$. This makes the qualitative difference between our work and previous studies.

\section*{II.   \ \ \ Fractional-wave molecule in spin-$1/2$ systems }

In this section, we provide details in constructing anyonic two-body states in spin-1/2($\uparrow,\downarrow$) systems. In the first part, we concentrate on spin-1/2 fermions with equal positive s- and p-wave scattering length $l_s=l_p>0$. This system can support s- and p-wave bound states with equal binding energy. In the second part, we briefly discuss two-species systems with hard-core s-wave interaction but no p-wave interaction, i.e., $l_s=l_p=0$. This system can support degenerate s- and p-wave states due to spin-charge separation in hard-core limit. 

\subsection*{A. \ \ \ $l_s=l_p>0$}

In the presence of positive scattering length $l_s=l_p>0$, both s- and p-wave molecules can be supported with the same binding energy $E_b$. Note that in the presence of finite effective range, it is still possible to tune both scattering length and effective range to ensure an equal binding energy for s- and p-wave molecules. In this way, their  real-space wavefunctions can be exactly mapped to each other. To simplify the discussions, we will assume zero effective range in the following. Moreover, here we will just consider spin-1/2 fermions, since  bononic system can easily form cluster states (beyond two-body correlations) under attractive interaction.

Consider two harmonically trapped fermions ($\uparrow,\downarrow$) with $l_s=l_p>0$, we can write down the wavefunctions of s- and p-wave molecules as Eqs.(9,10) in the main text. Explicitly, we have $\phi_p(x)=\epsilon(x)\phi_s(x)$ and 
\begin{equation}
\phi_s(x)=\frac{\cal{N}}{\sqrt{l_r}}\mathrm{e}^{- \frac{x^2}{2l_r^2}} \Gamma(-\nu) U\biggl(-\nu,\frac{1}{2}, \frac{x^2}{l_r^2}\biggr),
\end{equation}
where $\cal{N}$ is the normalization factor, $l_r=\sqrt{2}l_{ho}$ is the trap length for relative motion, $\Gamma$ is the Gamma function, $U$ is the confluent hypergeometric function and $\nu=\frac{E_b}{2\omega_{ho}}-\frac{1}{4}$, with $E_b$ determined by $2l/l_{r}=\Gamma(-\nu)/\Gamma(-\nu+1/2)$.

When a weak spin-orbit coupling (Eq.(11) in the main text) is acted on these molecules, we have
\begin{align}
&V_{soc}\Psi_{s}(R,x)=2\sqrt{2}\Omega 
(-i)\sin\frac{qx}{2} \left( e^{-i q R}\ket{\uparrow_1\uparrow_2}+e^{i q R}\ket{\downarrow_1\downarrow_2} \right) \Phi_0(R)\phi_s(x); \label{V_s}
\\
&V_{soc}\Psi_{p}(R,x)=2\sqrt{2}\Omega 
\cos\frac{qx}{2} \left( e^{-i q R}\ket{\uparrow_1\uparrow_2}+e^{i q R}\ket{\downarrow_1\downarrow_2} \right) \Phi_0(R)\phi_p(x). \label{V_p}
\end{align}
We can see that $V_{soc}$ changes the spin states and therefore it does not take effect in the  first-order perturbation. However, it  can give rise to an effective Hamiltonian ($H_{eff}$) in the second-order perturbation, in which process $\Psi_s$ and $\Psi_p$ are both virtually excited to $|\uparrow\uparrow\rangle$ and $|\downarrow\downarrow\rangle$ states at various levels of relative and CoM motions (note that the spatial wavefunctions of these relative motions should be anti-symmetric due to Fermi statistics). Up to a common factor, the coefficients in $2\times 2$ matrix  of $H_{eff}$ (see Eq.(12) in the main text) are
\begin{align}
&A=\sum_{N,n}\frac{1}{N+n+1/2-E_{b}/\omega_{ho}}\left(\int dx~\sin\frac{qx}{2} \phi_s(x) \phi_{n}^{(0)}(x) \right)^2 \ \left|\int dR~ e^{iqR}\Phi_{0}(R)\Phi_{N}^{(0)}(R) \right|^2; 
\\
&B= \sum_{N,n}\frac{1}{N+n+1/2-E_{b}/\omega_{ho}}\left(\int dx~\cos\frac{qx}{2} \phi_p(x) \phi_{n}^{(0)}(x) \right)^2\  \left|\int dR~ e^{iqR}\Phi_{0}(R)\Phi_{N}^{(0)}(R) \right|^2;
\\
&C=\sum_{N,n}\frac{1}{N+n+1/2-E_{b}/\omega_{ho}}\left( \int dx~\sin\frac{qx}{2} \phi_s(x) \phi_{n}^{(0)}(x)\right) \left( \int dx~\cos\frac{qx}{2} \phi_p(x) \phi_{n}^{(0)}(x)\right) \ \left|\int dR~ e^{iqR}\Phi_{0}(R)\Phi_{N}^{(0)}(R) \right|^2.
\end{align}
Here $\phi_{n}^{(0)}$ and $\Phi_{N}^{(0)}$ are, respectively, the non-interacting eigen-states of relative and CoM motions (note that $\Phi_0\equiv\Phi_{N=0}^{(0)}$); moreover, the index $n$ is an odd number in all above summations. 

From the expression of $C$, we can see that its sign changes with $q$. So when $q$ is converted to $-q$, we have $C\rightarrow-C$ and $\alpha \rightarrow -\alpha$. This offers a convenient control on the sign of $\alpha$.  

In the following we derive the momentum distributions of fractional-wave molecule. To facilitate the derivation, we assume the first atom is associated with spin-$\uparrow$  and the second atom is with spin-$\downarrow$, then the wavefunction (Eq.(13) in the main text) is reduced to  
\begin{equation}
\Psi_{G}(x_{1\uparrow},x_{2\downarrow})=\Phi_0(\frac{x_{1\uparrow}+x_{2\downarrow}}{2})\left[\phi_{s}(x_{2\downarrow}-x_{1\uparrow})-i\tan\frac{\alpha}{2}\ \phi_p(x_{2\downarrow}-x_{1\uparrow})\right]. \label{wf_spin}
\end{equation}
One can see that if $x_{2\downarrow}$ and $x_{1\uparrow}$ exchange, say, from $x_{2\downarrow}>x_{1\uparrow}$ to $x_{2\downarrow}<x_{1\uparrow}$, then the wavefunction will change by a phase factor  $e^{i\alpha}$, which directly manifests the fractional statistics characterized by  $\alpha$. Here we remark that this phase ($\alpha$) does not depend on the specific choice of spin state for the first or second atom. For instance, we can instead take the first atom as spin-$\downarrow$ ($x_{1\downarrow}$) and the second  as spin-$\uparrow$ ($x_{2\uparrow}$), then the wavefunction is similar to (\ref{wf_spin}) but with $i\rightarrow -i$ (according to Eq.(13) in the main text). In this case, we still get the same phase $\alpha$ if change the coordinates from $x_{1\downarrow}>x_{2\uparrow}$ to $x_{1\downarrow}<x_{2\uparrow}$. Therefore $\alpha$ can be uniquely determined by comparing the relative coordinates between spin-$\uparrow$ and spin-$\downarrow$, regardless of the labeling of each spin state. 

Based on (\ref{wf_spin}),  the momentum distributions of different spins are given by 
\begin{align}
n_{\uparrow}(q)&=\frac{1}{2\pi}\int dx_{1\uparrow} \int dx'_{1\uparrow} e^{iq(x_{1\uparrow}-x'_{1\uparrow})} \int dx_{2\downarrow} \Psi_{G}^*(x_{1\uparrow},x_{2\downarrow})\Psi_{G}(x'_{1\uparrow},x_{2\downarrow});
\\
n_{\downarrow}(q)&=\frac{1}{2\pi}\int dx_{2\downarrow} \int dx'_{2\downarrow} e^{iq(x_{2\downarrow}-x'_{2\downarrow})} \int dx_{1\uparrow} \Psi_{G}^*(x_{1\uparrow},x_{2\downarrow})\Psi_{G}(x_{1\uparrow},x'_{2\downarrow}).
\end{align}
After straightforward algebra, we have
\begin{align}
n_{\uparrow}(q)&=\int dk ~ \left|\Phi_{0}(K=2(q+k))\right|^2 \ \left| \phi_{s}(k)-i\tan(\frac{\alpha}{2})\phi_p(k) \right|^2;\\
n_{\downarrow}(q)&=\int dk ~ \left|\Phi_{0}(K=2(q-k))\right|^2 \ \left| \phi_{s}(k)-i\tan(\frac{\alpha}{2})\phi_p(k) \right|^2.
\end{align}
Here $\Phi_{0}(K),\ \phi_s(k),\ \phi_p(k)$ are the Fourier transformations of according real-space wavefunctions. From above expressions, we can see clearly that the system has the spin-momentum combined symmetry  $n_{\uparrow}(q)= n_{\downarrow}(-q)$ as long as the CoM part obeys reflection symmetry $\Phi_{0}(K)=\Phi_{0}(-K)$ (or $\Phi_{0}(R)=\Phi_{0}(-R)$). This can be satisfied in our case since $\Phi_0$ is the ground state of harmonic oscillators for CoM motion. 
For the continuum case where momentum is a good quantum number, we simply have $\Phi_{0}(K)=\delta(K)$ and therefore the momentum distributions will reduce to Eq.(17) in the main text.

\subsection*{B. \ \ \ $l_s=l_p=0$}

In the presence of zero scattering length $l_s=l_p=0$, $\uparrow$ and $\downarrow$ atoms have a hard-core s-wave interaction($U_s\sim -1/l_s\rightarrow \infty$) but no p-wave interaction (given the p-wave scattering amplitude proportional to $l_p$). With such a hard-core s-wave repulsion, spin and charge are well separated and different spin states (singlet or triplet) can have the same energy as long as their charge distributions are the same. This offers a natural situation for the coexistence of s- and p-wave states in Eq.(8) of the main text. 

Take two fermions ($\uparrow$ and $\downarrow$) in a harmonic trap for example, in the hard-core limit their charge part is equivalent to two free fermions described by a Slater determinant, leaving the spin part free (as singlet or triplet). Specifically, the two states, $\Psi_s$ and $\Psi_p$, are still as the form of Eqs.(9,10) in the main text, just with
\begin{equation}
\phi_p(x)=\epsilon(x)\phi_s(x)=\phi_1^{(0)}(x), \label{hard_core}
\end{equation}
with $\phi_1^{(0)}$ the first excited state ($n=1$) for the  relative motion. 

For two harmonically trapped bosons denoted by $\uparrow$ and $\downarrow$, in the hard-core limit the two states are
\begin{eqnarray}
\Psi_s(R,x)&=&\Phi_0(R)\phi_s(x)|\uparrow_1\downarrow_2+\downarrow_1\uparrow_2\rangle/\sqrt{2}; \\
\Psi_p(R,x)&=&\Phi_0(R)\phi_p(x)|\uparrow_1\downarrow_2-\downarrow_1\uparrow_2\rangle/\sqrt{2},
\end{eqnarray}
where $\phi_{s}$ and $\phi_p$ are the same as in Eq.(\ref{hard_core}). Note that an essential difference between fermions and bosons  is that the s-wave interacting bosons are ferromagnetic (spin-triplet in $\Psi_s$), while the s-wave interacting fermions are anti-ferromagnetic (spin-singlet in $\Psi_s$). This actually reflects the intrinsic relation between quantum statistics and interaction channels, as emphasized in the introduction part of the main text.

Under a small perturbative SOC, $\Psi_s$ and $\Psi_p$ can be transferred to other spin states ($\uparrow\uparrow$, $\downarrow\downarrow$), as shown similarly in Eqs.(\ref{V_s},\ref{V_p}). The specialty in this case ($l_s=l_p=0$) is that $\Psi_{s,p}$ here are all repulsive states, but not bound states as in last section ($l_s=l_p>0$). These repulsive states can have equal or even higher energies as compared to the eigen-states in $\uparrow\uparrow$ and $\downarrow\downarrow$ channels. Therefore, to well separate $\uparrow\downarrow$ from the other spin channels, an additional magnetic detuning ($\sim B \sigma_z$) can be applied. In this way one can effectively work with $\Psi_s$ and $\Psi_p$ basis within certain energy region, while treat other spin channels as virtual states. Then we can get the effective $2\times 2$ matrix expanded by $\Psi_s$ and $\Psi_p$, just as ${\cal M}$ matrix shown in Eq.(12) of the main text. The key point is that the off-diagonal elements are still purely imaginary, as seen from the consequence of SOC acting on $\Psi_{s,p}$ (see Eqs.(\ref{V_s},\ref{V_p}) for example). This guarantees the construction of anyonic two-body states as the form of Eq.(13) in the main text.


\begin{thebibliography}{99}
\bibitem{footnote} The definition of scattering length $l$ according to the behavior of $\Psi$ at $x\rightarrow 0^+$ (rather than $x\rightarrow 0^-$) is to ensure a two-body bound state supported at a positive (rather than negative) $l$.
\bibitem{footnote2} Note that $d^{\dag}$ itself is a bosonic operator, while the anyonic statistics comes out when exchanging constituent spins within a molecule.
\end{thebibliography}
\end{document}